\documentclass[aps,pra,twocolumn,groupedaddress,showpacs]{revtex4-1}
\usepackage{amsmath}
\usepackage{graphicx}

\usepackage{color}

\begin{document}

\title{Coupled square well model and Fano-phase correspondence}
\author{Bin Yan and Chris H. Greene}
\email{chgreene@purdue.edu}
\affiliation{
Department of Physics and Astronomy, Purdue University, 
West Lafayette, Indiana 47907, USA}
\date{\today}
\begin{abstract}
This paper investigates the Fano-Feshbach resonance with a two-channel coupled-square-well model in both the frequency and time domains. This systems is shown to exhibit Fano lineshape profiles in the energy  absorption spectrum. The associated time-dependent dipole response has a phase shift that has recently been understood to be related to the Fano lineshape asymmetric $q$ parameter by $\varphi=2\arg(q-i)$. The present study demonstrates that the phase-$q$ correspondence is general for any Fano resonance in the weak coupling regime, independent of the transition mechanism.
\end{abstract}


\maketitle

\section{Introduction}

Understanding time-dependent quantum dynamics has emerged as one of the fundamental problems in physics \cite{realtime}. In recent years, with the development of new technologies, especially with ultrashort light sources and ultrafast optical techniques, it has become possible to experimentally probe the real time electron dynamics in the quantum regime \cite{Pfeifer:16,Salieres:16,Huillier:16,Neumark:16,Taieb:14,Leonea:15,Lin:14,Fu:15}, e.g., time-domain measurements of the autoionization dynamics using attosecond pulses \cite{ZChang10,ZChang102,CDLin14}, creation and control of time-dependent electron wave packet \cite{Nisoli10,Pfeifer14-2}. Studying time domain resonance physics has been attracting increasing interests in atomic and molecular physics \cite{Jungen13,Burgdorfer13,Maquet14,Ivanov:15,Lin:10,Wickenhauser:05}.

A recent study \cite{Pfeifer13} has both theoretically established and experimentally verified a general correspondence between the photon absorption lineshape in the frequency domain, which is characterized by a Fano lineshape asymmetry parameter $q$ \cite{Fano61}, and the phase shift $\varphi$ of its time-dependent dipole response: 
\begin{equation}
\label{eq:phaseq}
\varphi=2\arg(q-i).
\end{equation}
In a further development, it was shown that by coupling the system with a short pulsed laser immediately after the excitation, the phase $\varphi$ of its dipole response can be externally controlled. In this way, the $q$-parameter of the system's subsequent absorption spectrum can be effectively modified. In the frequency domain, the Fano $q$ parameter provides a sensitive test of atomic structure calculations under field-free conditions \cite{Hibbert75, Chernoff11}. The phase-$q$ relation thus provides a possible way to control aspects of the time-dependent quantum dynamics.

The above infrared laser pulse control mechanism of the phase shift $\varphi$ was explained by both a quasi-classical, ponderomotive-motion picture \cite{Pfeifer13} and in terms of resonant coupling dynamics \cite{Pfeifer14}. However, the universal phase-$q$ correspondence Eq.(1) was only demonstrated as a general macroscopic property of a dielectric system, though it has already been faithfully applied to scenarios far beyond the area of atomic physics, e.g., condensed matter systems \cite{Tung:16,Mizoguchi:15}, plasma systems\cite{Gray:13,Walther:13}, high energy processes \cite{Ke:16}, or optomechanics systems \cite{Saif:15}. Thus, it would be more interesting to have a unified treatment of this phase-$q$ relation for any Fano resonance.

In the present work, with all these questions in mind, we focus on an analytically solvable two-channel square well model and study its resonance physics in both the frequency and time domains. As an extension of a textbook single-channel square-well scattering problem, the coupled-channels model captures much of the physics of near-threshold bound and scattering states \cite{Chin10}. This model has been used to successfully explain the threshold scattering of cold neutrons from atomic nuclei \cite{Bethe35} and to represent Feshbach resonances in ultracold atom scattering processes \cite{Kokkelmans02,Duine04,Chin:09}. Investigation of the Fano-phase correspondence with such a model would then generalize the previous result in the dielectric atomic systems to a more general class of scenarios, and thus extend its potential applications.

This paper is organized as following. Section \ref{sec2} introduces the two-channel coupled-square-well model. By adding an auxiliary ground state belonging to a third channel, we study the energy absorption spectrum when the system is excited from the ground state to the coupled two channels through magnetic dipole transitions. A standard Fano lineshape is observed for the absorption cross section, with the asymmetry $q$ parameter linearly depending on the transition dipole ratio $d_2/d_1$, consistent with Fano's configuration interaction theory \cite{Fano61}. In section \ref{sec3}, the dipole response in the time domain is studied. The phase-$q$ correspondence Eq. (\ref{eq:phaseq}) is revealed numerically in the present model problem for magnetic dipole transitions. A general proof of this relation for any transition mechanism is also presented. Finally, Sec. \ref{sec4} summarizes our conclusions. Derivation of the eigen solutions and discussions of the scattering properties of the two-channel square-well model are given in the Appendix.

\section{Coupled Square Well Model}\label{sec2}

The two-channel square well model in the present study describes two particles with reduced mass $m$ interacting in three dimensions with the following $s$-wave Hamiltonian in the relative coordinate $r$:
\begin{equation}
\label{eq:modelhami}
\hat{H}=-\frac{\hbar^2}{2m}\hat{I}\frac{d^2}{dr^2}+\hat{V}(r)+\hat{E}^{th}.
\end{equation}
Here the potential coupling matrix is assumed to vanish at $r > r_0$, but a constant $2\times2$ matrix at $r < r_0$: 
\begin{equation}
\hat{V}(r)= 
\begin{bmatrix}
-V_1 & V_{12}\\
V_{12}& -V_2
\end{bmatrix}
\theta(r-r_0).
\end{equation}
We are most interested in the case for which the diagonal elements are attractive, which is why a negative sign has been separated out from this equation at the outset, given that $V_1$ and $V_2$ are positive. The matrix $\hat{E}^{th}$ containing the real energy thresholds is diagonal. We choose the lower threshold, channel $|1\rangle$ in our notation, as defining the zero of our total energy scale, whereby
\begin{equation}
\hat{E}^{th}=
\begin{bmatrix}
0&0\\0&E_2^{th}
\end{bmatrix}.
\end{equation}
This model has a single analytic solution between the two energy thresholds, which contains both an exponentially decaying solution in the closed channel $|1\rangle$ and a scattering solution in the open channel $|2\rangle$. In matrix form the channel bases read $|1\rangle=\begin{bmatrix}1\\0\end{bmatrix}$ and $|2\rangle=\begin{bmatrix}0\\1\end{bmatrix}$. The radial parts of the energy eigenfunctions are linear combinations of the two channels' configuration basis functions: $|\epsilon\rangle=\phi_1(r;\epsilon)|1\rangle+\phi_2(r;\epsilon)|2\rangle$, the derivation of which is presented in the Appendix.

For appropriate potential parameters, there exist one or more bound states below the lower energy threshold. Up to this point, external field excited transition from these bound states can then be investigated. However, in order to simplify the model without losing the key features of the problem, while making it extendable to problems involving more than two channels, we model the ground state with an auxiliary channel $|0\rangle$ independent with channel $|1\rangle$ and $|2\rangle$. This might correspond to an independent degree of freedom in realistic systems, such as hyperfine spin state in a cold atom pair.

Suppose initially the system is prepared in the s-wave ground state $|g\rangle=f(r)|0\rangle$, where $f(r)$ is the normalized radial part of the ground state wave function in the coordinate representation. In this paper, we consider the ground state to be strongly localized, and for definiteness we take $f(r)=2 e^{-2r}$. 
At $t=0$, a strong $\delta$ pulse couples the auxiliary channel $|0\rangle$ to the two channels through magnetic dipole interaction, and then consequently excites the system to channel $|1\rangle$ and $|2\rangle$. The short pulse in modeled by a delta interaction,
\begin{equation}
\hat{H}_\delta=\gamma\delta(t)\hat{d}+h.c.,
\end{equation}
where $\gamma$ is a dimensionless interaction strength parameter. $\hat{d}=d_1|1\rangle\langle0|+d_2|2\rangle\langle0|$ is the transition dipole. Parameters $d_1$ and $d_2$ control the transition strengths into the corresponding channels. \\

\begin{figure}
\includegraphics[width=8.5cm]{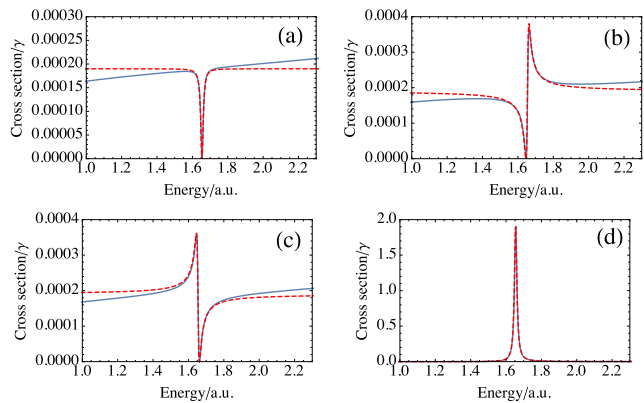}
\caption{\label{fig:CroSec} Resonance profiles at various transition dipoles are showns as cross sections versus the energy. The solid blue lines are the numerically calculated absorption cross sections. Dashed red lines are standard Fano lineshapes (Eq.(\ref{eq:fanocros})) with a background cross section $\sigma_0$=1.9E-4, resonance position $\epsilon_r$=1.65 a.u. and resonance width $\Gamma=1.7E-2$.  Model potential parameters are fixed at $V_1=75, V_2=10, V_{12}=10, E^{th}_2=2, r_0=3$. The transition dipole parameter $d_1$ is fixed to be 1. The corresponding Fano lineshape parameters and values of $d_2$ are a) $d_2=-0.137, q=0$; b) $d_2=-0.130, q=1$; c) $d_2=-0.144, q=-1$ and d) $d_2=0.600, q=100$.}
\end{figure}

The wavefunction immediately after the excitation pulse is given by
\begin{equation}
\label{eq:initialwf}
\begin{aligned}
|\psi(t=0)\rangle &= e^{-i\int_{0_-}^{0_+} dt \hat{H}_\delta t} \,|g\rangle \\
&= e^{-i\gamma\hat{d}}\,|g\rangle.
\end{aligned}
\end{equation}
In the perturbative limit, where $\gamma\ll 1$, the evolution operator is expanded to the first order of $\gamma$:
\begin{equation}
\begin{aligned}
|\psi(t=0)\rangle&\approx |g\rangle-i\gamma\hat{d}\,|g\rangle \\
&= |g\rangle-i\gamma\int d\epsilon \, \langle\epsilon|\hat{d}|g\rangle |\epsilon\rangle,
\end{aligned}
\end{equation}
where $|\epsilon\rangle$ denotes the energy eigenstates. The photo absorption cross section can then be calculated as
\begin{equation}
\begin{aligned}
\sigma(\epsilon) &=|\langle\psi|\epsilon\rangle|^2\\
&=\gamma^2|\langle\epsilon|\hat{d}|g\rangle|^2.
\end{aligned}
\end{equation}

\begin{figure}
\includegraphics[width=8cm]{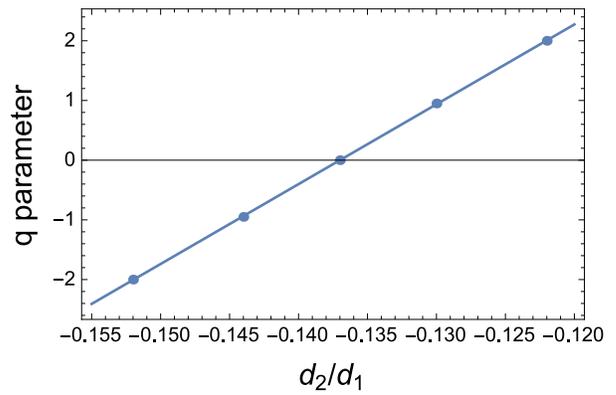}
\caption{\label{fig:qvd}$q$ versus $d_2/d_1$. Dots are numerically fitted $q$ parameters, which exhibit a linear dependence on $d_2/d_1$, as expected from Eq. (\ref{eq:qpara})}
\end{figure}

According to Fano's configuration interaction theory, in the energy range between the two threshold energies, where the bound states in the first channel are coupled to the continuum of the second channel, the resonance profile at each resonance point is predicted to have a simple form:

\begin{equation}
\label{eq:fanocros}
\sigma_{Fano}(\epsilon)=\sigma_{0}\frac{(q+\bar{\epsilon})^2}{1+\bar{\epsilon}^2},
\end{equation}
where $\bar{\epsilon}=\frac{\epsilon-\epsilon_r}{\Gamma/2}$.
With the assumption of a flat-background near resonance and constant coupling potential $V$, the q parameter is defined by
\begin{equation}
q \equiv \frac{\langle\alpha|\hat{d}|g\rangle}{\pi V \langle\beta_E|\hat{d}|g\rangle},
\end{equation}
where $|\alpha\rangle$ and $|\beta_E\rangle$ are respectively the bare closed-channel bound state and the unperturbed open-channel energy-normalized continuum eigenstate. In the present model problem, for fixed system potential parameters, it can be further deduced that
\begin{equation}
\label{eq:qpara}
q \propto \frac{d2}{d1}.
\end{equation}
In our numerical study, we tune the potential parameters such that there is exactly one bound state in the first channel, and such that the background cross section is relatively flat near the position of the resonance. In Fig. (\ref{fig:CroSec}) the cross sections for different transition probabilities are plotted. The Fano and Lorentz line profiles can both be realized by tuning the ratio between $d_2$ and $d_1$. Fig. (\ref{fig:qvd}) shows the numerically fitted $q$ parameters at various values of $d_2/d_1$, which matches the linear relation as predicted by Eq. (\ref{eq:qpara}).

\section{Dipole Response} \label{sec3}

With the aid of the time dependent wavefunction after the strong short pulse excitation,
\begin{equation}
\label{eq:wavefunction}
|\psi(t)\rangle = e^{-i\epsilon_g t}|g\rangle-i\gamma\int d\epsilon \, e^{-i\epsilon t} \langle\epsilon|\hat{d}|g\rangle |\epsilon\rangle,
\end{equation}
the dipole response function can be calculated as the quantum average of the dipole transition operator:
\begin{equation}
\begin{aligned}
d(t) &= \langle\psi(t)|\hat{d}|\psi(t)\rangle \\
&= 2 \mathit{Re}[i\gamma\int d\epsilon \, \langle g|\hat{d}^\dag|\epsilon\rangle\langle\epsilon|\hat{d}|g\rangle e^{i(\epsilon-\epsilon_g) t}] \\
&= \frac{2}{\gamma} \mathit{Im}[\int d\epsilon \, \sigma(\epsilon) e^{-i(\epsilon-\epsilon_g) t}],
\end{aligned}
\end{equation}
where $\mathit{Re}$ and $\mathit{Im}$ denote the real part and imaginary part, respectively. In the case of a Fano resonance with cross section Eq.(\ref{eq:fanocros}), the above shifted Fourier transform can be evaluated directly:
\begin{equation}
\label{eq:TDdipole}
\begin{split}
\int d\epsilon \, \sigma_{Fano}(\epsilon) e^{-i(\epsilon-\epsilon_g) t} =  2\pi \sigma_0 \delta(t)\\
+\pi \sigma_0 (\Gamma/2)e^{-\frac{\Gamma}{2}t}e^{-i(\epsilon_r-\epsilon_g)t}(q-i)^2,
\end{split}
\end{equation}
giving the transition stength a delta function response, which comes from the non-zero background cross section $\sigma_0$, followed by a decaying single mode oscillation. 
The complex factor $(q-i)^2$ can be cast into exponential representation, $(q-i)^2=(q^2+1)exp[i\phi(q)]$, where $\phi(q)=2\arg(q-i)$ induces a phase shift. 

We note the following remarks: 1) The above derivation involves only the physically measurable real quantities $d(t)$ and $\sigma(\epsilon)$, and is general for any transition interaction and model Hamiltonian, as long as wavefunction Eq.(\ref{eq:wavefunction}) is valid in the perturbative limit. This generalizes the application of the phase-q correspondence Eq.(\ref{eq:phaseq}), which was originally developed in Ref. \cite{Pfeifer13} for macroscopic dielectric systems; in that study, the complex dipole in the energy domain $\tilde{d}(\epsilon)$ and relation $\sigma(\epsilon)\propto \mathit{Im}[\tilde{d}(\epsilon)]$ \cite{FanoCooper68} were used as the starting point. 2) The frequency of the dipole response is of course the transition energy between the ground state and the resonance energy (neglecting the small resonance level shift due to discrete-continuum level mixing), in agreement with Ref.\cite{Pfeifer13} including the phase shift and $q$ parameter relationship. In the following numerical study of the phase shift, we always ignore the ground state energy, i.e., setting $\epsilon_g=0$ \footnote{Strictly speaking, the ground state energy cannot be zero as the lower energy threshold has already been defined to be zero. However, since the ground state energy only changes the overall oscillation frequency but not the phase information, for simplicity we compute the dipole response with $\epsilon_g=0$, which is equivalent as changing the time scale to $(\epsilon_r-\epsilon_g)/\epsilon_r$.}.

\begin{figure}
\includegraphics[width=8.5cm]{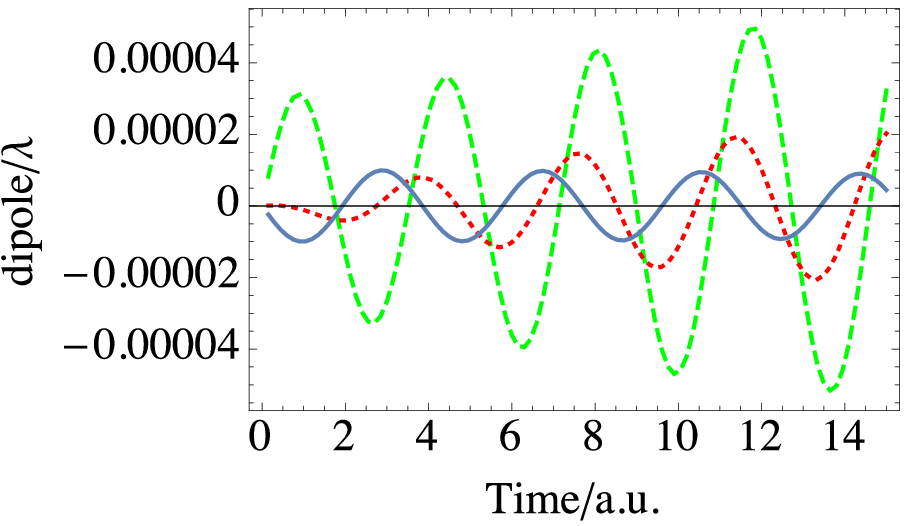}
\caption{\label{fig:dipole} Time-dependent dipole response for different q-parameters. The model potential parameters are chosen as in Fig. \ref{fig:CroSec}. The solid blue line, dotted red line and dashed green line correspond to $q=0, 1, 2$ respectively.}
\end{figure}

\begin{figure}
\includegraphics[width=8.5cm]{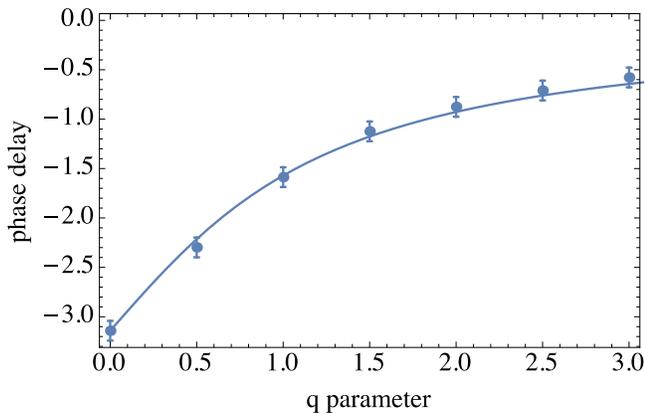}
\caption{\label{fig:phaseq}Phase shift versus $q$ parameters. Dots are values read out from the explicit numerical calculation of the time dependent dipole. The solid line is the $q$-phase relation of Eq. (\ref{eq:phaseq}).}
\end{figure}

The physically measurable dipole response function Eq.(\ref{eq:TDdipole}) is understood to be exact when the integral is ranging over the whole range of the energy spectrum, in which case the dipole response would be a complicated mixture of different frequencies. When one is interested in observing the effect of an isolated resonance at position $\epsilon_r$, namely, the particular frequency mode $(\epsilon_r-\epsilon_g)/2\pi$ of the dipole response, the integral can be restricted to a finite range of a few resonance widths near the resonance energy, e.g., $[\epsilon_r-\Delta E,\epsilon_r+\Delta E]$. However, in this manner the dipole response would be dependent on the choice of $\Delta E$ - the integral in Eq.(\ref{eq:TDdipole}) does not converge with $\Delta E$ because of the non-zero background cross section. To eliminate this dependence without changing the critical phase information, in the following numerical study we compute the dipole response using the shifted cross section $\sigma(\epsilon)-\sigma_0$.
In Fig. \ref{fig:dipole}, the time-dependent dipole response for different Fano $q$-parameters are plotted. The phase shifts are read out and compared with the phase-$q$ correspondence Eq. (\ref{eq:phaseq}) in Fig. (\ref{fig:phaseq}).

The above analysis shows the decay of dipole response and the change of line shapes corresponding to different transition parameters, i.e., tuning of the internal system parameter. On the other hand, it is more interesting to show that for a system with fixed internal parameters, resonance profiles can be modified through controlling with external field. Similar to the treatment in Ref. \cite{Pfeifer13}, we introduce a subsequent control pulse, applied to the system immediately after the first excitation pulse, modeled by the following interaction:
\begin{equation}
\hat{H}_2=\beta e^{i\phi}\delta(t)\hat{D},
\end{equation}
where $\hat{D}=d_1|1\rangle\langle1|+d_2|2\rangle\langle2|$ is the magnetic dipole moment or more generally the transition operator. The frequency of the second pulse is assumed to be far from resonance such that excitation of the ground state channel by the control pulse would not take place. In the region that the control pulse is much shorter than the lifetime of the system, we treat the pulse as a $\delta$ function, with two free parameters left to be tuned: the strength $\beta$ characterizing the overall effect of the intensity and duration of the pulse, and the phase shift $\phi$, characterizing the phase offset between the initial time of the system evolution and the control pulse when interaction is turned on. To see the effect of the second pulse, we apply the evolution operator to the initial state (\ref{eq:initialwf}):
\begin{equation}
\begin{aligned}
|\psi'(t=0)\rangle&=e^{-i\beta e^{i\phi}\hat{D}}|\psi(t=0)\rangle\\
&\approx (1-i\beta e^{i\phi}\hat{D})(1-i\gamma\hat{d})|g\rangle \\
&= |g\rangle-i\gamma(\hat{d}-i\beta e^{i\phi}\hat{D}\hat{d})|g\rangle,
\end{aligned}
\end{equation}
which, compared with state (\ref{eq:initialwf}), has an overall effect of modifying the original transition dipole operator $\hat{d}$. For appropriate tuned phase shift, e.g., $\phi=\pm\pi/2$, the effective transition dipole becomes
\begin{equation}
\hat{d}\to d_1(1\pm\beta d_1)|1\rangle\langle0|+d_2(1\pm\beta d_2)|2\rangle\langle0|,
\end{equation} 
or
\begin{equation}
\frac{d_1}{d_2}\to\frac{d_1}{d_2}(1+\beta(d_1\mp d_2)).
\end{equation}
Combined with the fact that the Fano line shape parameter is proportional to the ratio between $d_2$ to $d_1$, it is  concluded that the control pulse will lead to an effective change of $q$-parameter, namely, a modification of the Fano line shape.

\section{Conclusion} \label{sec4}

To summarize, we have investigated the Fano resonance with an analytic solvable coupled-square-well model in both frequency and time domain. The Fano asymmetric parameter $q$ and the phase shift $\phi$ of the magnetic dipole transition were shown to have a simple relation $\psi=2\arg(q-i)$, which generalizes the result originally discovered in Ref.\cite{Pfeifer13} for electric dipoles. This relation was also proven to be valid for any transition dipole, as long as an isolated Fano resonance is present in the perturbative limit.

\begin{acknowledgments}
This work was supported by the U.S. Department of Energy, Office of Science, under Award No. DE-SC0010545.
\end{acknowledgments}

\appendix

\section{Solution of the model}
We present here the derivation of the solution of the energy eigenfunctions and discuss the scattering behavior of the coupled two square well model. 
The time independent Schrodinger equation of Hamiltonian (\ref{eq:modelhami}) would possess 4 independent solutions in general, but when we restrict the solutions to obey the regular physical constrains $\psi(r=0)=0$, only two linearly independent solutions remain. 
Our solution strategy will be to begin by solving for these two
linearly-independent solutions that are regular at the origin.  In a second
step, we will impose long-range boundary conditions, enforcing exponential delay in the closed channel, and determine the exact
S-matrix for this model so we can study its poles in the complex energy
plane.  Since the coupling potential is constant within the reaction
volume, it can be diagonalized by an $r$-independent eigenvector matrix,
which reduces the solution to two uncoupled short-range eigenchannels.
First, define the matrix 

\begin{equation}
\begin{aligned}
\hat{W}=&\left( -\frac{2m}{\hbar ^{2}}\right) \left( \hat{V}+%
\hat{E}^{th}-\varepsilon \hat{I}\right) \\
=&\left( \frac{2m}{\hbar
^{2}}\right) \left( 
\begin{array}{cc}
\varepsilon +V_{1} & -V_{12} \\ 
-V_{12} & \varepsilon -E_{2}^{th}+V_{2}%
\end{array}
\right) ,
\end{aligned}
\end{equation}
and indicate the constant orthogonal eigenvector matrix as $X_{i\alpha }$
and the (weakly) energy-dependent diagonal eigenvalue matrix by $w_{\alpha}(\varepsilon )^{2}.$  Thus we have, in matrix notation, $\hat{W}\hat{X}=\hat{X}\hat{w}^{\text{2}}$. Next we replace the
solution matrix $\hat{u}(r)$ by $\hat{X}\hat{X}^{T}\hat{u}(r)$
just before the solution matrix $\hat{u}(r)$ in the time-independent
Schrodinger equation $\hat{u}^{\prime \prime }(r)+\hat{W}\hat{u}(r)=0$.  Upon left-multiplying the whole equation by $\hat{X}^{T}$, we obtain two uncoupled single-channel equations in the
eigenrepresentation. The diagonal eigensolution matrix at short range will
be denoted $\hat{y}(r)=\hat{X}^{T}\hat{u}(r)$, and the components of this solution obey the 2nd order equation, $y_{\alpha
}^{\prime \prime }(r)+w_{\alpha }{}^{2}y_{\alpha }(r)=0$. The regular
solution at the origin is of course $\sin \left( w_{\alpha }r\right)$. 

The next step consists of matching this solution to the simple trigonometric
solutions that apply outside the reaction volume, at $r>r_{0}$, and imposing
the physically relevant boundary conditions at $r\rightarrow \infty$. The
correct physical solution at all distances $r>r_{0}$ is of course a
scattering solution in the open channel $|1\rangle$, and an exponentially decaying
solution in the closed channel $|2\rangle$: 
\begin{equation}
\vec{\psi}^{phys}(r)=\left( 
\begin{array}{c}
e^{ikr}S-e^{-ikr} \\ 
Ne^{-qr}%
\end{array}
\right) .
\end{equation}
Here $S$ is the desired scattering matrix at energy $\varepsilon$, while $N$
is a closed-channel amplitude, which we organize into a column vector $\vec{s}=\left( 
\begin{array}{c}
S \\ 
N%
\end{array}
\right)$. The components of this vector will be determined by matching this form for the outer region solution and derivative to our short range solution derived above, at $r=r_{0}$.  Note that $k^{2}=2m\varepsilon /\hbar ^{2}$, while $q^{2}=2m(E_{2}^{th}-\varepsilon )/\hbar ^{2}$. Neither of our two
short-range eigensolutions will in general match smoothly onto this desired
long range behavior. We must superpose the two solutions with constant
coefficients $\vec{z}=\{z_{1},z_{2}\}^{T}$\ in order to accomplish this. This leads to a set of continuity equations with the structure: 
\begin{eqnarray}
\vec{\psi}^{phys}(r_{0}) &=&\hat{X}\hat{y}(r_{0})\vec{z}=-\vec{a}%
(r_{0})+\hat{D}(r_{0})\vec{s} \\
\vec{\psi}^{phys\;\prime }(r_{0}) &=&\hat{X}\hat{y}^{\prime
}(r_{0})\vec{z}=-\vec{a}^{\prime }(r_{0})+\hat{D}^{\prime }(r_{0})\vec{%
s}.
\end{eqnarray}
\newline
Here, for notational convenience, we have defined a vector 
\begin{equation}
\vec{a}(r)=\left( 
\begin{array}{c}
e^{-ikr} \\ 
0%
\end{array}
\right)
\end{equation}
and a diagonal matrix 
\begin{equation}
\hat{D}(r)=\left( 
\begin{array}{cc}
e^{ikr} & 0 \\ 
0 & e^{-qr}%
\end{array}
\right).
\end{equation}

The next step is to eliminate $\vec{z}=-\hat{y}(r_0)^{-1}\hat{X}^{T}\vec{a}(r_{0})+\hat{y}(r_{0})^{-1}\hat{X}^{T}\hat{D}(r_{0})\vec{s}$, and insert it into the derivative continuity equation, giving 

\begin{equation}
\begin{aligned}
&\hat{X}\hat{y}^{\prime }(r_{0})[-\hat{y}(r_{0})^{-1}\hat{X}^{T}\vec{a}(r_{0})+\hat{y}(r_{0})^{-1}\hat{X}^{T}\hat{D}(r_{0})\vec{s}]\\
=&-\vec{a}^{\prime }(r_{0})+\hat{D}^{\prime }(r_{0})\vec{s}.
\end{aligned}
\end{equation}
Wigner's real, symmetric R-matrix is now evident in this equation, and it
will simplify our algebra if we denote it explicitly:

\begin{equation}
\begin{aligned}
\hat{R}=&\hat{X}\hat{y}^{\prime }(r_0)\hat{y}(r_0)^{-1}\hat{X}^{T}\\
=&\hat{X}\left( 
\begin{array}{cc}
w_{1}\cot w_{1}r_{0} & 0 \\ 
0 & w_{2}\cot w_{2}r_{0}%
\end{array}
\right) \hat{X}^{T}.
\end{aligned}
\end{equation}

The above equation now reads $-\hat{R}\vec{a}(r_0)+\hat{R}\hat{D}%
(r_{0})\vec{s}=-\vec{a}^{\prime }(r_{0})+\hat{D}^{\prime }(r_{0})\vec{s%
}$. Thus we obtain our solution for the physically important quantities
contained in $\vec{s}$:

\begin{equation}
\vec{s}=\left( \hat{D}^{\prime }(r_{0})-\hat{R}\hat{D}(r_{0})\right)
^{-1}\left( \vec{a}^{\prime }(r_{0})-\hat{R}\vec{a}(r_{0})\right) .
\end{equation}

More explicitly, 

\begin{equation}
\begin{split}
\left( 
\begin{array}{c}
S \\ 
N%
\end{array}
\right) =\left( 
\begin{array}{cc}
\left( ik-R_{11}\right) e^{ikr_{0}} & -R_{12}e^{-qr_{0}} \\ 
-R_{21}e^{ikr_{0}} & \left( -q-R_{22}\right) e^{-qr_{0}}%
\end{array}
\right) ^{-1}\\
\left( 
\begin{array}{c}
-ik-R_{11} \\ 
-R_{21}%
\end{array}
\right) e^{-ikr_{0}}
\end{split}
\end{equation}

Now the scattering matrix $S$ is readily evaluated, but instead of giving that explicit formula here, we give instead the formula for the poles of $S$. These occur at energies for which

\begin{equation}
(q+R_{22})(-ik+R_{11})-R_{12}^2=0.
\end{equation}

This equation could now be solved numerically to determine the pole positions in the complex energy plane. However, it will be consistent with the other approximations we have made to this point if we make a linear expansion of $q$ about zero energy and about the magnetic field point $B_0$ at which a new bound state appears or disappears. Our approximate treatment will neglect the energy and field dependences of the R-matrix itself, and assume that the closed channel wavenumber $q$ depends on energy as is evident in its definition above, and on magnetic field through an assumed variation of the upper threshold energy with magnetic field, i.e. $E^{th}_2 = E^{th}_2(B)$, whereby we can write

\begin{equation}
q(\epsilon,B)\simeq q_0+\zeta k^2+\gamma(B-B_0).
\end{equation}
Here the two real constants $\zeta$ and $\gamma$ are defined by
\begin{equation}
\begin{aligned}
&\zeta\equiv\frac{\hbar^2}{2m}[\frac{\partial q(\epsilon,B)}{\partial\epsilon}]_{\epsilon=0,B=B_0},\\
&\gamma\equiv[\frac{\partial q(\epsilon,B)}{\partial B}]_{\epsilon=0,B=B_0}.
\end{aligned}
\end{equation}
Three pole locations now emerge as the roots of a cubic equation in $k$, at any chosen field value $B$. The
fact that the scattering length at $k = 0$ is infinite when $B = B_0$ implies that $q_0$ is fixed to have the value
$q_0=-(R_{11}R_{22}-R^2_{12})/R_{11}$, which brings our final cubic equation to the form:
\begin{equation}
ik\frac{R^2_{12}}{\zeta R_{11}}+(R_{11}-ik)(\frac{\gamma}{\zeta}B^{'}+k^2)=0.
\end{equation}
Interestingly, there are 3 real parameters that control the structure of these S-matrix poles in the complex energy plane, namely $R_{12}$ ,$R_{11}$, and $\gamma/\zeta$. Each of these can be assigned a direct physical interpretation in this problem. First of all, $R_{11}$ can be approximately associated with the background scattering length, i.e. $Y\equiv R_{11} \simeq -1/A_{bg}$, provided $A_{bg} \gg r_0$, as is usually the case for the atom-atom s-wave scattering in most alkali systems. Notice that $Z \equiv\frac{\hbar^2\gamma}{2m\zeta}$ is the slope of the Feshbach resonance, i.e. the variation of the resonance energy per unit change of the magnetic field. Finally, the parameter $X\simeq R^2_{12}/\zeta R_{11}$ is a measure of the coupling strength between the channels, giving
\begin{equation}
ikX+(Y-ik)(ZB^{'}+k^2)=0.
\end{equation}

The actual scattering amplitude itself takes the following form, in terms of the original R-matrix elements:
\begin{equation}
S=e^{-2ikr_0}\frac{R^2_{12}-R_{11}R_{22}-qR_{11}-ik(q+R_{22})}{R^2_{12}-R_{11}R_{22}-qR_{11}+ik(q+R_{22})}.
\end{equation}

In thinking about the energy dependence of this scattering matrix, it should be remembered that each element of the R-matrix is in general a meromorphic function of the energy. However, since the scale of short-range interactions is typically huge compared to the ultra-cold energy scale, it will usually be a good approximation to regard each element of the R-matrix as energy-independent in applications at sub-microkelvin temperatures. Also, a linear expansion of $q$ as a function of $\epsilon$ and $B$ can be inserted, as was discussed above.

It may be interesting to contrast this expression with the exact single-channel result for a short-range potential. The most general S-matrix for the single channel problem has the form:
\begin{equation}
S=e^{-2ikr_0}\frac{R(\epsilon)+ik}{R(\epsilon)-ik}.
\end{equation}
Here again, the most general energy-dependence for $R(\epsilon)$ is a meromorphic function with poles on the real energy axis.

\section{Physical scattering length}

The physical scattering length of the system at zero energy can be extracted from the low-energy behavior of the scattering phase shift. First, however, it is useful to define the (weakly) energy-dependent scattering length, in terms of the exact S-wave scattering phase shift:
\begin{equation}
a(\epsilon, B)\equiv -\frac{tan\delta(\epsilon, B)}{k}.
\end{equation}
The zero-energy scattering length is typically used in the context of BECs and DFGs, which is of course just the zero energy limit of this last expression, or in terms of the scattering amplitude derived earlier, 
\begin{equation}
a(0, B)=\lim_{\epsilon\to 0}(-\frac{1}{2ik} \ln S(\epsilon, B)).
\end{equation}
This gives the following for the zero-energy scattering length as a function of magnetic field:
\begin{equation}
\begin{aligned}
a(0, B)&\simeq\frac{R^2_{12}+\gamma R_{11}(B-B_0)}{-\gamma R^2_{11}(B-B_0)}+r_0\\
&\equiv a_{bg}(1-\frac{\Delta}{B-B_0}).
\end{aligned}
\end{equation}
This expression is valid only at zero energy, and over the range of magnetic field values for which $q$ can be expanded linearly in $B \approx B_0$. The next important correction term should be included when $B \approx B_0$, where the denominator should include a linear function of energy in order to obtain a more general and effective parameterization, i.e.
\begin{equation}
a(\epsilon, B)\simeq a_{bg}(1-\frac{\Delta}{B-B_0+\zeta\epsilon}).
\end{equation}
This form for the general phase shift is very accurate, typically within approximately 1 to 10 microkelvin above and below zero energy.

\section{Bound State Energy Level Properties}

The above wavefunction used to describe low energy atom-atom scattering still applies at negative energies, $\epsilon=\frac{-\hbar^2\kappa^2}{2m}$. For definiteness, I assume that the analytic continuation in going from positive to negative energies is carried out by setting $k\to i\kappa$, with the convention that $\kappa$ is a real, positive number in this regime. Then the entire derivation could be repeated from the beginning, of course, but a shorter route to the desired result just begins from the above unnormalized wavefunction, except we divide it by $S(\epsilon, B)$ . The wavefunction in the ?weakly-closed? channel is then
\begin{equation}
\psi\to e^{-\kappa r}-e^{\kappa r}S^{-1},
\end{equation}
which will be unphysical and diverge exponentially unless $S^{-1} \to 0$ for some $\kappa>0$. Referring to the above expression
for $S$, the condition for a bound state thus becomes:
\begin{equation}
\kappa=\frac{R^2_{12}-R_{11}R_{22}-qR_{11}}{q+R_{22}}.
\end{equation}
The linear expansion can now be inserted, $q \simeq q_0 + \gamma(B-B_0)$, i.e. neglecting the weak energy dependence of $q$. When this result is combined with the fact that the point at which the scattering length is infinite has been defined to be $B_0$, the bound state wavenumber is seen to be given simply by:
\begin{equation}
\kappa=\frac{-\gamma R^2_{11}(B-B_0)}{R^2_{12}+\gamma R_{11}(B-B_0)}.
\end{equation}
Since the bound state energy is $\epsilon=\frac{-\hbar^2\kappa^2}{2m}$ (provided the preceding expression for $\kappa$ is positive), this proves that the binding energy of a high-lying bound level always approaches 0 quadratically in the magnetic field, except in the uninteresting limit where the channels are noninteracting.

Another quantity of physical interest is the probability that the system resides in the upper (strongly-closed)
channel. In the limit of a zero-range potential $r_0 \to 0$, and in the limit of very small binding where $\kappa\to1/A$, this probability is given by
\begin{equation}
Probability(|2\rangle)\simeq\frac{(1+AR_{11})^2}{(1+AR_{11})^2+A^3q_0R^2_{12}},
\end{equation}
which vanishes as $1/A$ in the limit where the physical scattering length diverges, i.e. when $B \to B_0$ and $A \to\infty$.

\bibliography{main}

\end{document}